\documentclass[epj]{webofc}
\usepackage[varg]{txfonts}   
\wocname{EPJ Web of Conferences}
\woctitle{CONF12}
\begin{document}
\selectlanguage{english}
\title{Nuclear physics insights for new-physics searches using nuclei:\\
Neutrinoless $\beta\beta$ decay and dark matter direct detection}

\author{Javier Men\'{e}ndez\inst{1}
\fnsep\thanks{\email{menendez@nt.phys.s.u-tokyo.ac.jp}}
}

\institute{Department of Physics, The University of Tokyo,
Hongo, Bunkyo-ku, Tokyo 113-0033, Japan 
}

\abstract{%
Experiments using nuclei to probe new physics beyond the Standard Model,
such as neutrinoless $\beta\beta$ decay searches
testing whether neutrinos are their own antiparticle,
and direct detection experiments aiming to identify the nature of dark matter,
require accurate nuclear physics input
for optimizing their discovery potential
and for a correct interpretation of their results.
This demands a detailed knowledge of the nuclear structure
relevant for these processes.
For instance, neutrinoless $\beta\beta$ decay nuclear matrix elements
are very sensitive to the nuclear correlations in the initial and final nuclei,
and the spin-dependent nuclear structure factors of dark matter scattering
depend on the subtle distribution of the nuclear spin among all nucleons.
In addition, nucleons are composite and strongly interacting,
which implies that many-nucleon processes are necessary
for a correct description of nuclei and their interactions.
It is thus crucial that theoretical studies and experimental analyses
consider $\beta$ decays and dark matter interactions
with a coupling to two nucleons, called two-nucleon currents.
}
\maketitle
\section{Introduction}
\label{intro}

Neutrinos and dark matter are two of the most promising candidates
for new physics beyond the Standard Model of particle physics.
Because they are both charge neutral and massive,
it is possible that neutrinos and antineutrinos would be the same particle,
in which case neutrinos would be labeled
as Majorana particles~\cite{Avignone08}.
This property ---very hard to test because neutrinos are so light---
would imply the violation of lepton number,
a relation which goes in both directions:
the violation of lepton number would establish neutrinos
to be Majorana particles.
In turn, this may have important consequences for the understanding of
the baryonic matter-antimatter asymmetry observed in the universe,
as in most models the difference between baryon and lepton number is conserved.

Unveiling the origin of dark matter
stands as one of the biggest challenges in physics.
The existence of dark matter has been certified
by very different astrophysical observations
---galactic rotation velocities,
gravitational lensing, anysotropies of the cosmic microwave background---
but the nature of dark matter is still unknown~\cite{Baudis:2016qwx}.
Observations have constrained some of its properties,
such that it must be neutral to the electromagnetic interaction
---to a very good approximation at least---
that it should be cold or warm to allow for galaxy structure formation,
and that it amounts to more than 80\% of the mass content of the universe,
and roughly a quarter of the energy content.

Ideally one would like to answer these questions on the nature of neutrinos
and dark matter in the laboratory.
Experimental programs searching for neutrinoless $\beta\beta$ decay
---the lepton number violating process most likely to be observed---
and the direct detection of dark matter are being pursued vigorously,
and impressive advances are permanently being reported,
with present experimental sensitivities reaching half-lives longer than
$T_{1/2}^{0\nu\beta\beta}=10^{26}$~years~\cite{KamLAND-Zen:2016pfg}
for neutrinoless $\beta\beta$ decay,
and excluding scattering cross-sections off nuclei smaller than
$\sigma_{\chi\mathcal{N}}=10^{-40}$~cm$^2$~\cite{Akerib:2015rjg}
for searches of dark matter.
Further improvements are expected in the near future,
as next generation experiments are planned to use
over a tonne of source or target material
with increasingly reduced backgrounds.

Neutrinoless $\beta\beta$ decay and dark matter direct detection experiments
have in common that they are looking for the decay and the scattering off
atomic nuclei, respectively.
Therefore, the design ---for example, the choice of source or target material---
and the interpretation of the experimental results
in principle depends on the nuclear physics of the process at study.
In the case of neutrinoless $\beta\beta$ decay,
the value of the nuclear matrix element driving the transition
relies on the accurate nuclear structure description of the initial and final
nuclei, and on the weak-interaction diagrams considered at the nucleon level
\cite{Engel:2016xgb}.
In the case of dark matter detection,
a correct interpretation of the experimental results taking into account
all relevant nuclear structure factors depends on considering all
possible interactions of nuclei with dark matter particles.
In particular, $\beta$ decays and dark matter interactions
with a coupling to two nucleons,
in addition to the leading contributions which only involve a single nucleon,
can be significant. 

\section{Neutrinoless $\beta\beta$ decay}

\subsection{$\beta\beta$ decay: two-neutrino and neutrinoless cases}

\begin{figure}[b]
\centering
\sidecaption
\includegraphics[width=9cm,clip]{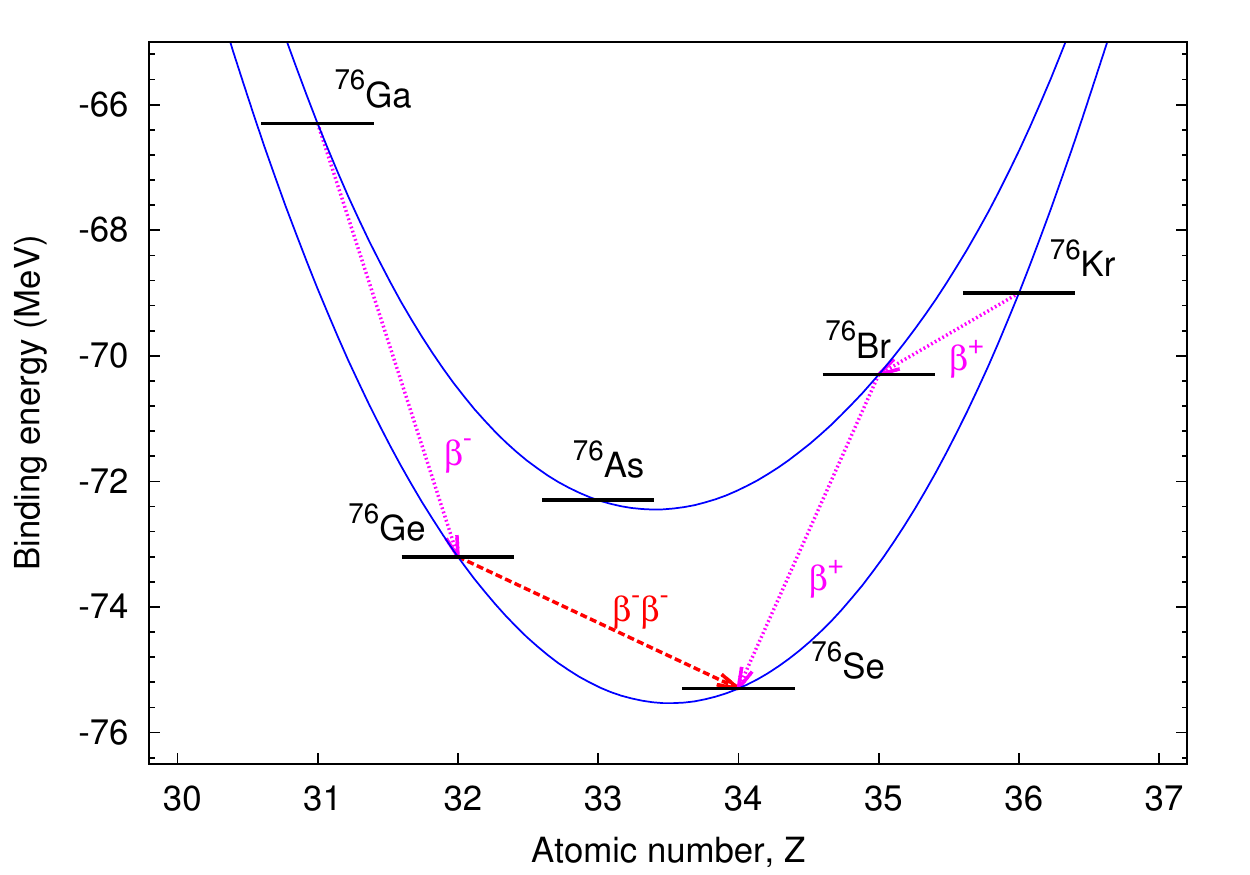}
\caption{Nuclear binding energies and decay scheme
of the nucleon number $A=76$ isobars, as a function of the atomic number $Z$.
In a typical situation, isotopes decay to the lowest-energy nucleus
via single $\beta^-$ or $\beta^+$ decays. For the special case of $^{76}$Ge,
however, single-$\beta$ decay is energetically forbidden,
leaving $\beta\beta$ decay as the only channel
available to reach the stable nucleus $^{76}$Se.}
\label{f:bb_decay}       
\end{figure}

The existence of $\beta\beta$ decay
is a consequence of the nuclear pairing interaction,
which makes nuclei with an even number of protons,
or an even number of neutrons, more bound than nuclei with one or two
---a proton and a neutron--- unpaired nucleons.
As a result, in some cases it is energetically favorable
for a nucleus to decay along a given isobaric chain
---set of nuclei with the same number of nucleons---
via a second-order $\beta\beta$ decay,
instead of the usual single-$\beta$ decay channel. 
For the case of $^{76}$Ge the decay scheme is shown in figure~\ref{f:bb_decay}.
$\beta\beta$ decay with the emission of two antineutrinos besides two electrons,
a lepton-conserving process permitted by the weak interaction,
has been observed in a dozen of cases, favored by a larger energy
difference between the initial and final nuclei.
The measured half-lives are of the order of
$T_{1/2}^{2\nu\beta\beta}\sim10^{19}-10^{21}$~years~\cite{Barabash15}.

Neutrinoless $\beta\beta$ decay does not involve the emission of neutrinos,
and it therefore violates lepton number.
It demands neutrinos to be Majorana particles.
The neutrinoless case is at least five or six orders of magnitude slower
than the two-neutrino $\beta\beta$ decay permitted by the Standard Model.
In the standard scenario where the decay
is mediated by the exchange of the three known light neutrinos
this is because the decay rate is proportional to the neutrinos masses,
which are tiny compared to those of any other lepton.
In other scenarios involving new physics,
the reason is the large mass of the exchange particles,
or the small coupling of the new physics with the Standard Model sector.
In the standard case the neutrinoless $\beta\beta$ decay half-life
can be written as~\cite{Avignone08}
\begin{equation}
\label{eq:finalrate}
\left[T^{0\nu\beta\beta}_{1/2}\right]^{-1}=G^{0\nu\beta\beta}
  \left|M^{0\nu\beta\beta}\right|^2  m_{\beta\beta}^2~,
\end{equation}
which naturally includes a phase space factor $G^{0\nu\beta\beta}$
that takes into account the kinematics,
a nuclear matrix element $M^{0\nu\beta\beta}$
that contains the relevant nuclear physics of the decay,
and a third part $m_{\beta\beta}=\sum_k m_kU_{ek}^2$
that encodes the new-physics scale ---the neutrino masses $m_k$---
and also includes the mixing of electron neutrinos with other flavors, $U_{ek}$.
The nuclear matrix element can be decomposed
according to the spin structure of the operator~\cite{Avignone08}:
\begin{equation}
M^{0\nu\beta\beta}=M^{GT}-\frac{g_V}{g_A}M^{F}+M^{T},
\end{equation}
where the dominant term is the so-called Gamow-Teller component, $M^{GT}$.

\subsection{Nuclear matrix elements: nuclear structure}

\begin{figure}[t]
\centering
\includegraphics[width=9.5cm,clip]{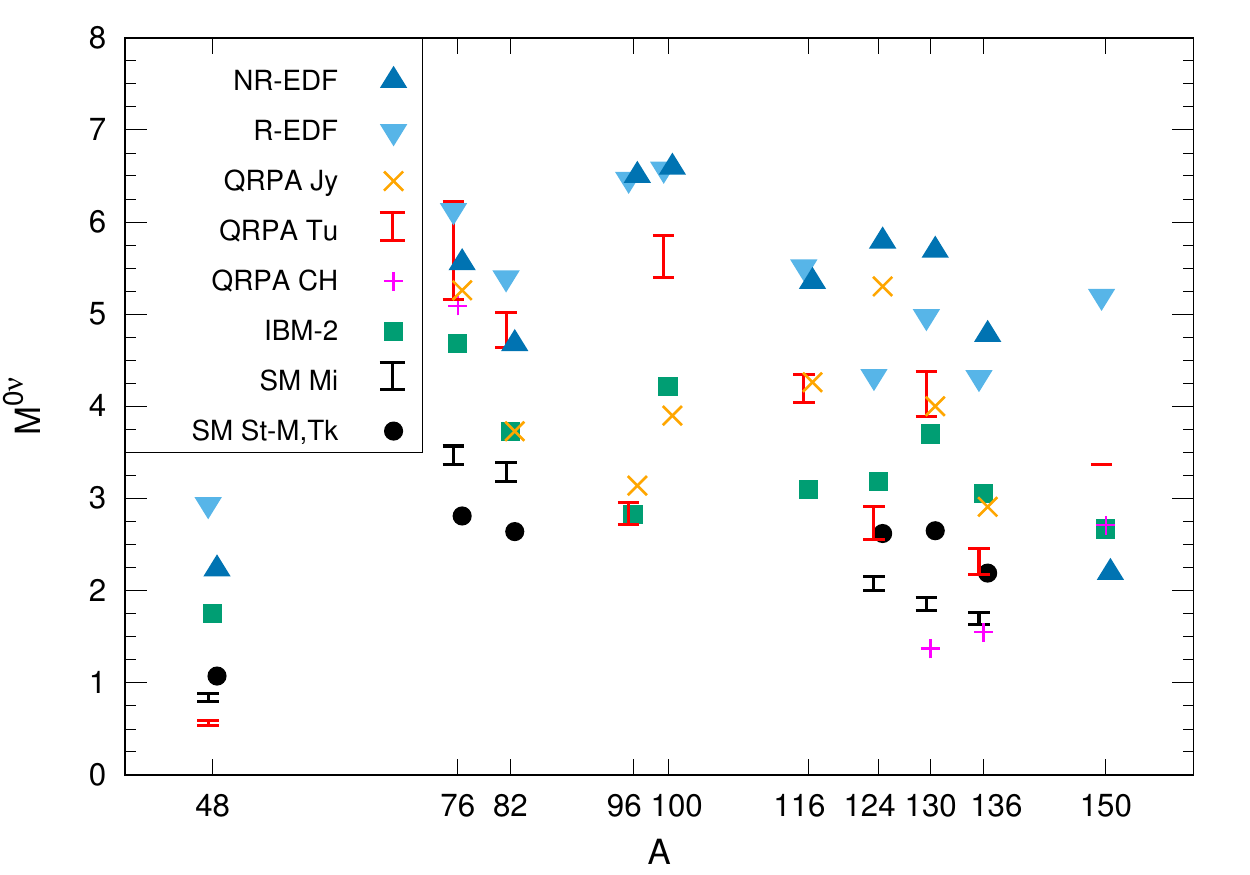}
\caption{Nuclear matrix elements $M^{0\nu\beta\beta}$
for neutrinoless $\beta\beta$ decay candidates,
shown as a function of the nucleon number $A$,
calculated with different nuclear structure methods.
Blue symbols indicate energy density functional (EDF) calculations,
nonrelativistic~\cite{Vaquero13} (triangles)
and relativistic~\cite{Yao15,Yao16} (down triangles).
Orange crosses, red bars and magenta plus signs stand for
quasiparticle random phase approximation (QRPA) results obtained by
the Jyv{\"a}skyl{\"a}~\cite{Hyvarinen15}, T{\"u}bingen~\cite{Simkovic13,Fang15}
and Chapel Hill~\cite{Mustonen13} groups, respectively.
Green squares show interacting boson model (IBM) values~\cite{Barea15}.
Finally, black bars and circles represent the shell model (SM) results
obtained by the Michigan~\cite{Horoi16}
and Strasbourg-Madrid~\cite{Menendez09} groups, respectively,
except for the $^{48}$Ca black circle which shows
the recent shell model calculation by the Tokyo group~\cite{Iwata16}.}
\label{f:nme}       
\end{figure}

Neutrinoless $\beta\beta$ decay nuclear matrix elements $M^{0\nu\beta\beta}$
have to be obtained by nuclear structure calculations
evaluating the transition operator between the initial and final nuclear states.
The present status of these calculations is illustrated in figure~\ref{f:nme}.
Unfortunately different nuclear structure approaches disagree
in their predicted matrix elements for every $\beta\beta$ decay candidate
by up to a factor three.
This is a clear evidence that the unavoidable approximations present
in solving the nuclear many-body problem
are not under control when studying neutrinoless $\beta\beta$ decay
\cite{Engel:2016xgb}.
In contrast, it should be noted that the same different many-body methods
in general agree when studying other nuclear structure properties 
such as excitation spectra or electromagnetic transitions.

It is thus critical to clarify the actual value
of neutrinoless $\beta\beta$ decay nuclear matrix elements.
A first avenue for doing so is to test the calculations
by finding correlations of matrix elements with other measured quantities.
Despite efforts in this direction~\cite{Freeman12},
no single observable has been found to be especially correlated
to neutrinoless $\beta\beta$ decay.
A promising process is two-neutrino $\beta\beta$ decay,
which shares with the neutrinoless case the initial and final states
and for which there is experimental data.
It must be taken into account, however,
that the relevant momentum transfers are very different
in the two-neutrino and neutrinoless decays, $q\sim1$~MeV for the former
and $q\sim100$~MeV for the latter.
Unfortunately most many-body approaches
cannot predict two-neutrino $\beta\beta$ decay,
because an accurate calculation involves dealing with
the intermediate odd-proton--odd-neutron system
---for instance, the nucleus $^{76}$Sb in figure~\ref{f:bb_decay}---
and this is a more involved nuclear structure calculation
than the one needed for even-even nuclei.
Other approaches like quasiparticle random phase approximation method
use the two-neutrino $\beta\beta$ decay half-life
to fix a free parameter in their model and are not predictive for this decay.
The only remaining many-body approach is the shell model,
which actually predicted the two-neutrino $\beta\beta$ decay rate of $^{48}$Ca
\cite{Caurier90}
in good agreement with the subsequent measurement a few years later
\cite{Balysh96}.
However, the accepted experimental value has been challenged
by a recent measurement~\cite{Nemo316},
which would bring the shell model prediction
into an overestimation of the corresponding matrix element.
The description of the two-neutrino $\beta\beta$ decay rate of $^{136}$Xe
within the shell model is also under discussion~\cite{Caurier12,Horoi13}.

\begin{figure}[t]
\centering
\sidecaption
\includegraphics[width=9.5cm,clip]{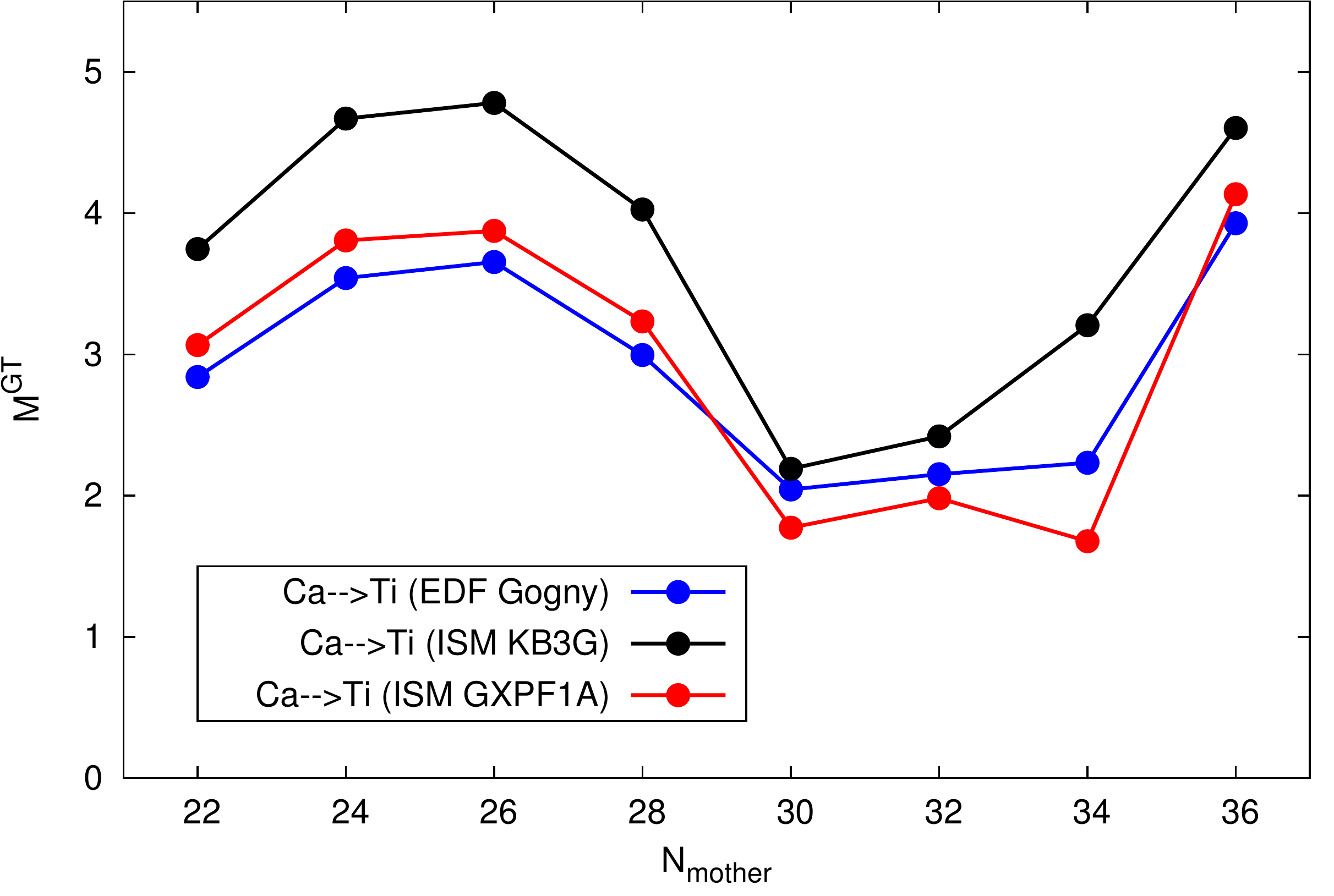}
\caption{Gamow-Teller component
of the neutrinoless $\beta\beta$ decay matrix element, $M^{GT}$,
for the decay of calcium into titanium isotopes,
as a function of the neutron number of the calciums $N_{\text{mother}}$
\cite{Menendez14}.
Results from an energy density functional calculation (blue)
are compared to shell model results
obtained with two different effective interactions (black and red).}
\label{f:nocorrel}
\end{figure}

In recent years an effort has been made to understand
the origin of the differences between many-body approaches
by comparing the systematic calculations of matrix elements,
even if the associated $\beta\beta$ decays
make little or no sense experimentally.
For instance a comparison between shell model
and energy density functional matrix elements
which in general disagree the most between different methods,
as shown in figure~\ref{f:nme}, but restricting the calculations
to uncorrelated states ---fully composed of proton-proton and neutron-neutron angular momentum $J=0$ pairs in the shell model,
and limited to spherical states in the energy density functional calculation---
showed that the uncorrelated nuclear matrix element disagreement
was limited to 30\% or less~\cite{Menendez14}.
This is illustrated by figure~\ref{f:nocorrel}
and is a very significant improvement over the factor three disagreement
in figure~\ref{f:nme}.
The matrix element values are fixed by the strength
of the ---neutron-neutron and proton-proton--- pairing interaction.

This finding suggests that it is the different way
in which the many-body approaches include nuclear structure correlations
which is behind much of the disagreement
in neutrinoless $\beta\beta$ decay nuclear matrix elements.
Proton-neutron pairing has been known to be important for
$\beta$ and $\beta\beta$ decay decays for a long time~\cite{Vogel86}.
Recent shell model calculations based on a separable effective interaction
have confirmed this extreme, showing that nuclear matrix elements
are overestimated if proton-neutron pairing
---more precisely, isoscalar pairing--- is excluded~\cite{Menendez16}.
Since at the moment isoscalar pairing is not fully captured by
energy density functional and interacting boson model calculations,
this may be a reason for the discrepancies shown in figure~\ref{f:nme}.
Dedicated studies incorporating these correlations in the corresponding
nuclear matrix element calculations are needed.

Another cause of missing correlations
is the limitation of the configuration space, this is,
the single-particle orbitals that nucleons are permitted to occupy.
For instance, the shell model and the interacting boson model
only solve explicitly the nuclear many-body problem
in a limited configuration space around the Fermi surface,
and include the effect of the remaining configurations approximately.
In order to quantify the effect of the missing correlations,
a many-body perturbation theory estimate found relatively moderate increases
of less than 50\% for the lightest $\beta\beta$ decay emitters
$^{48}$Ca, $^{76}$Ge and $^{82}$Se, for which the estimation is easier
\cite{Holt13,Kwiatkowski14}.

A more rigorous estimation is the recent computation
of the $\beta\beta$ decay of $^{48}$Ca
extending the shell model configuration space
from one to two major harmonic oscillator shells
---limited to $2\hbar\omega$ excitations---
reducing the core of the shell model calculation from $^{40}$Ca to $^{16}$O.
By doing so, many previously excluded configurations were permitted,
and the dimension of the problem increases from less than $10^6$ to over $10^9$.
The impact on the nuclear matrix element is, however, moderate,
with a $30\%$ enhancement
due to previously-missing cross-shell pairing correlations~\cite{Iwata16}.
Interestingly, a cancellation occurs
between the general enhancement produced by additional pairing correlations
and the contribution of particle-hole excitations.
A similar cancellation is expected to be at play
for other $\beta\beta$ decay candidates as well.
Nevertheless, explicit calculations are needed.
Shell model calculations in extended configuration spaces
can benefit from using the Monte Carlo shell model technique,
which recently has studied configuration spaces with dimension over $10^{23}$
\cite{Togashi16}.

Finally, other many-body approaches than those represented in figure~\ref{f:nme}
can shed light into $\beta\beta$ decay nuclear matrix elements.
In particular, in the last decade nuclear ab initio calculations
---those solving the many-body problem for {\it all} nucleons in the system,
with nuclear forces fitted only to light nuclei---
have been able to perform calculations up to medium-mass isotopes,
in many cases achieving very good agreement to experiment~\cite{Hebeler15}.
Even before ab initio calculations are available
for $\beta\beta$ decay emitters, these many-body techniques
can be used to benchmark $\beta\beta$ decay matrix elements
in lighter or less-correlated systems to gain insight on the relevant physics
for this process. 
Not only ab initio calculations are more controlled
than the phenomenological ones available so far for $\beta\beta$ decay, 
but they also in principle allow
for the estimation of theoretical nuclear matrix element uncertainties,
a very valuable information for the interpretation of the experimental results
\cite{Engel:2016xgb}.

\subsection{Two-body currents}

\begin{figure}[t]
\centering
\includegraphics[width=9.5cm,clip]{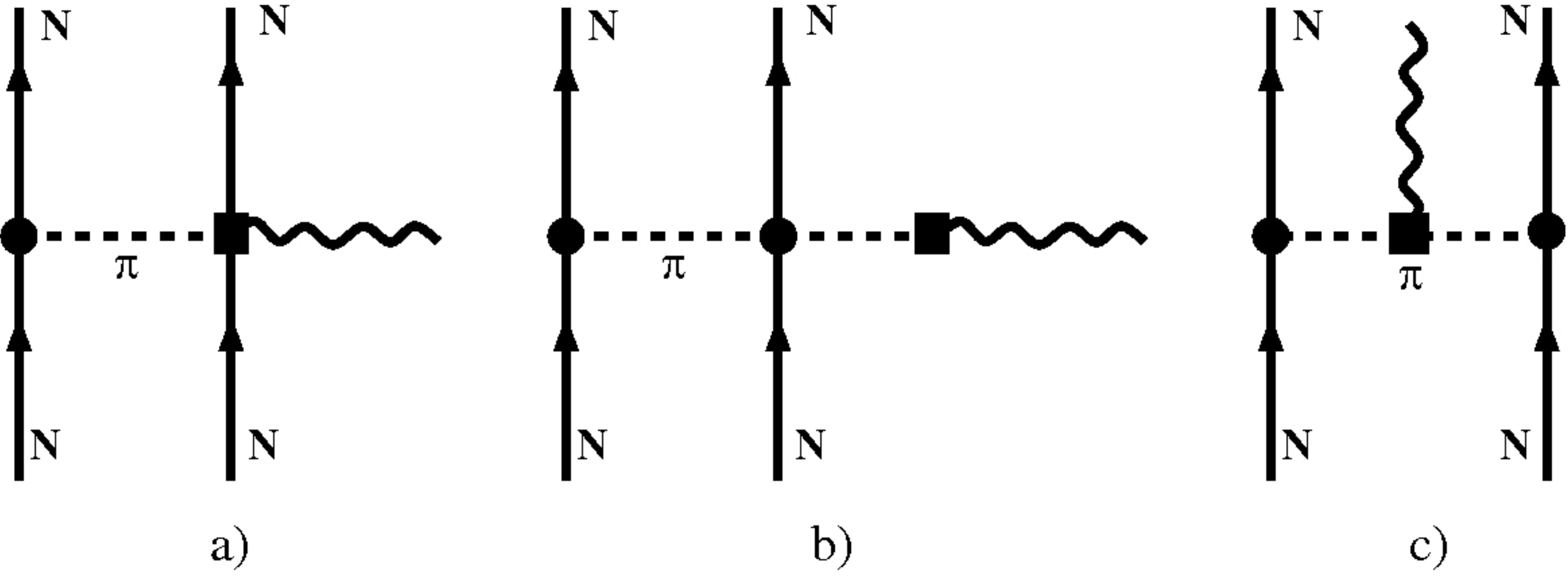}
\caption{Leading long-range two-body currents
contributing to neutrinoless $\beta\beta$ decay
\cite{MGS2011,Hoferichter:2015ipa,Menendez:2016kkg}
and dark matter scattering off nuclei
\cite{Prezeau:2003sv,Cirigliano:2012pq,Menendez:2012tm,Hoferichter:2016nvd}.
Solid and dashed lines indicate nucleons ($N$) and pions ($\pi$), respectively,
and wavy lines stand for the external source,
in this work weak and dark matter interactions that
can be approximated as contact couplings.
Diagrams a) and b) stand for one-pion exchange
and one-pion exchange pion-pole couplings
appearing in the leading axial two-body current.
Diagram c) represents the pion-in-flight coupling
in the leading vector and scalar two-body currents.
All three diagrams enter in both processes.
}
\label{f:2b_long} 
\end{figure}

The nuclear matrix discussed so far are based on the standard one-body operator:
each weak interaction vertex involves only one nucleon.
However, because nucleons are composite particles
that strongly interact to each other, the single-$\beta$ decay hadronic current 
takes the general form~\cite{Park:2002yp}
\begin{equation}
{\bf J}=\sum_{i=1}{\bf J}_{i,{\rm 1b}}+\sum_{i<j}{\bf J}_{ij,\text{2b}} +\cdots
\label{eq:1b2b3b}
\end{equation}
Diagrams involving two nucleons, represented in figure~\ref{f:2b_long},
are a correction to the leading one-body terms.
Two-body corrections have been intensively explored in nuclei lighter
than those undergoing $\beta\beta$ decay,
where they have been found to be relevant
for electromagnetic~\cite{Bacca:2014tla}
and weak~\cite{Gazit09} transitions.

Two-body weak currents have been mostly studied in the context
of small momentum transfers, such as single-$\beta$ decay.
Since calculations are quite demanding, most works
have been limited to relatively light nuclei,
up to about oxygen, or relatively rough approximations,
such as the normal-ordering of the two-body currents
with respect to an isospin-symmetric Fermi gas.
These studies indicate that two-body currents tend to cancel
the contribution of the leading one-body operator,
which means that they are a contribution to the so-called "$g_A$ quenching".
The "$g_A $ quenching" stands for the empirical fact that
nuclear many-body calculations need to reduce the strength
of the spin-isospin operator in order to agree with
the experimental half-lives of Gamow-Teller $\beta$ transitions.
However, the size of the two-body contributions is unclear,
with results ranging for less than 10\% for carbon and oxygen isotopes
\cite{Ekstrom:2014iya},
to about 30\% in larger systems but with a more rough normal ordering
\cite{MGS2011}.
Especially important are the uncertainties in the
short-range two-body currents
---not shown in figure~\ref{f:2b_long}.
The "$g_A$ quenching" should be carefully studied
in lighter systems where dedicated
ab initio calculations with different approaches are feasible.

In contrast to single-$\beta$ decay,
where the relevant momentum transfer is $q\sim1$~MeV,
in neutrinoless $\beta\beta$ decay transferred momenta can reach $q\sim100$~MeV,
because of the virtual nature of the exchanged neutrinos.
This different momentum-transfer regime can have important consequences
in the effect of two-body currents,
because several pion-exchange (figure~\ref{f:2b_long} a)
and pion-pole (figure~\ref{f:2b_long} b) contributions contribute at finite $q$
\cite{Hoferichter:2015ipa}.
Present results show that $q$-dependent two-body contributions
partially cancel other two-body terms~\cite{MGS2011,eng14,Menendez:2016kkg},
resulting in a smaller reduction of Gamow-Teller matrix elements
in neutrinoless $\beta\beta$ decay than in single-$\beta$
and two-neutrino $\beta\beta$ decays.

\section{Dark matter scattering off nuclei}

\subsection{Dark matter-nucleon interactions and scattering cross-section}

Direct detection dark matter searches are motivated by
weakly interacting massive particles (WIMPs), promising dark matter constituents
that are predicted to naturally account for the observed dark matter density.
The expected WIMP masses are $M_{\chi}\sim1-1000$~GeV,
the mass scale of nuclei.
Therefore experiments sensitive to these dark matter masses
use nuclei as target.
Similarly lighter masses $M_{\chi}\sim1$~MeV
are probed via the scattering of dark matter off electrons
~\cite{Baudis:2016qwx}.

WIMPs can interact with nuclei in many ways.
However, commonly the interactions suppressed by the small WIMP velocities
---$v/c\sim10^{-3}$--- or the momentum transfer of the scattering
---much smaller than $M_{\chi}$ and the nucleon mass--- are not considered.
This leads to two possibilities~\cite{Engel:1992bf}:
the direct coupling of the dark matter and nuclear densities,
called spin-independent scattering,
and the coupling of the dark matter and nuclear spins,
referred to as spin-dependent scattering.
Spin-independent scattering is favored from the nuclear physics side
because it is {\it coherent}: at vanishing momentum transfer
it receives contributions from all nucleons in the nucleus.
In contrast, in spin-dependent responses,
on average only the spin of one nucleon contributes,
because the spins of two nucleons tend to couple to a spin-zero pair
due to the pairing interaction.
Other things being equal, coherent scattering is expected to be enhanced
by a factor $A^2$ with respect to the spin-dependent case.

A more complete description of the possible WIMP interactions 
with nuclei has been worked out
in a nonrelativistic effective field theory (EFT)
\cite{Fitzpatrick:2012ix,Anand:2013yka}.
By constructing all possible interactions that can be built from
the WIMP and nucleon spins, the momentum transfer and
the WIMPs relative velocity,
a set of independent operators $\mathcal{O}_i$ is derived at the nucleon level.
At the nuclear level, however,
not all operators $\mathcal{O}_i$ leave distinct signatures,
because there are only six independent nuclear responses,
which may interfere between them.
In addition, the interactions of dark matter with two nucleons
need to be considered~\cite{Prezeau:2003sv,Cirigliano:2012pq,Menendez:2012tm},
see figure~\ref{f:2b_long}.
Two-body currents can contribute significantly to dark matter scattering,
both for a coherent response, and for spin-dependent interactions.

In general, the WIMP-nucleus cross-section
can be written as~\cite{Hoferichter:2016nvd}
\begin{align}
\frac{\text{d} \sigma_{\chi\mathcal{N}}}{\text{d} \mathbf{q}^2}
\propto \Big|\sum_i c_i \,\zeta_i\, \mathcal{F}_i \Big|^2
+\Big|\sum_i \hat{c}_i \,\hat{\zeta}_i\,\hat{\mathcal{F}}_i \Big|^2+\cdots
\label{eq:cross-section_gen}
\end{align}
Here $\zeta,\hat{\zeta}$ are kinematic factors,
$c,\hat{c}$ encode the hadronic physics and particle physics
---for instance the Wilson coefficients coupling WIMPs with quarks and gluons---
and $\mathcal{F},\hat{\mathcal{F}}$
represent the nuclear physics:
its square gives the nuclear structure factors.
As shown in Eq.~\eqref{eq:cross-section_gen},
contributions may interfere or not.
In particular, the usual spin-independent
and spin-dependent terms do not interfere.

\subsection{Coherent (spin-independent) scattering}

Spin-independent ---coherent--- scattering can be generalized
by considering all one-nucleon operators proposed by the nonrelativistic EFT.
The most relevant terms
are characterized by an enhancement of the cross-section, 
driven by the nuclear physics,
and reflected in the structure factors $\mathcal{F}^2,\hat{\mathcal{F}}^2$.

In the nonrelativistic EFT there are two
nuclear responses which can be coherent~\cite{Fitzpatrick:2012ix,Anand:2013yka}.
First, the standard spin-independent response, denoted as $M$,
corresponding to the operator $\mathcal{O}_1$ and other subleading operators.
Second, the nuclear response associated with the operator $\mathcal{O}_3$,
denoted by $\Phi''$, which is partially coherent.
In this case, all nucleons with spin aligned with the angular momentum $l$ 
contribute coherently.
Nucleons with spin antiparallel to $l$ cancel this contribution,
but single-particle states with parallel spin are lowered in energy
due to the nuclear spin-orbit force, leaving part
of the antiparallel-spin states empty and preventing a complete cancellation.
Since this coherent term is spin dependent,
the standard terminology must be generalized: it is more appropriate
to speak of coherent scattering instead of spin-independent scattering.

Besides the nuclear structure aspects ---coherence---
the hadronic physics is also crucial to set the hierarchy among different terms.
Chiral EFT~\cite{Epelbaum:2008ga,Machleidt:2011zz},
an effective theory of the underlying interaction that binds nucleons,
quantum chromodynamics (QCD), is valid at the energy and momentum scales
of WIMP scattering and incorporates the physics of the chiral symmetry of QCD.
Chiral EFT also describes consistently the interactions
with external probes~\cite{Bacca:2014tla}.
By formulating the WIMP-nucleon interactions
in the chiral EFT framework, including scalar, pseudoscalar, vector and axial
contributions in the WIMP and hadronic sectors,
the leading operators are predicted and can be matched
into the nonrelativistic EFT basis~\cite{Hoferichter:2015ipa}.
In addition, chiral EFT predicts the consistent interactions
of WIMPs with two nucleons ---two-body currents.
The relative importance of the different contributions
can be studied by calculating the corresponding structure factors
for the one- and two-body operators,
assuming similar contributions from the particle physics Wilson coefficients.

\begin{figure}[t]
\centering
\includegraphics[width=13.5cm,clip]{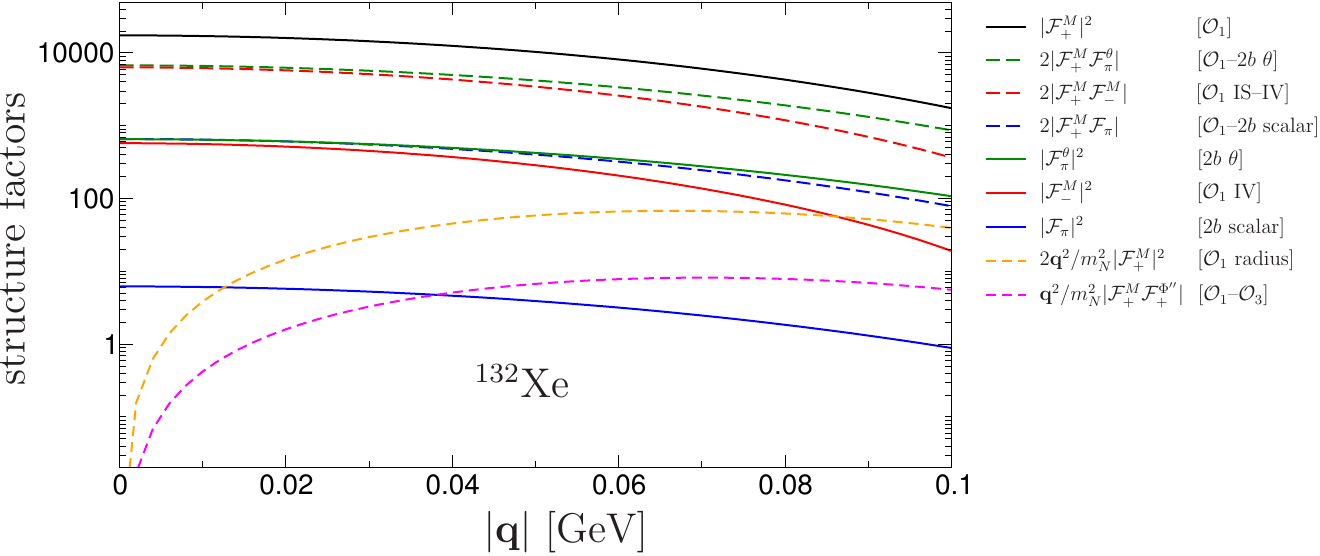}
\caption{Structure factors associated with different WIMP interactions
with $^{132}$Xe, as a function of the momentum transfer $q$
\cite{Hoferichter:2016nvd}.
The solid (red) black and line shows
the isoscalar (isovector) $\mathcal{O}_1$ contributions.
The green and blue solid lines denote the two-body current contributions
with scalar and $\theta$-term couplings, respectively,
while the dashed lines with the same colors stand for their interference
with the isoscalar $\mathcal{O}_1$ term.
Orange (magenta) dashed lines denote the interference
of the one-nucleon radius corrections ($\mathcal{O}_3$ operator)
with the $\mathcal{O}_1$ term.}
\label{f:structure_factors} 
\end{figure}

Figure~\ref{f:structure_factors} shows the structure factors
for the most important one- and two-nucleon contributions to the coherent
WIMP scattering off $^{132}$Xe.
This is the most abundant isotope in xenon, the target
used in experiments giving the present best limits on WIMP-nucleus scattering.
Solid lines in figure~\ref{f:structure_factors} represent
the most important individual contributions,
which are the isoscalar $\mathcal{O}_1$ term
---routinely used in experimental analyses---
its isovector counterpart ---with opposite coupling to protons and neutrons,
usually included to account for apparently conflicting results
in experiments using different isotopes---
and two different couplings of the WIMP to two nucleons:
through a scalar coupling,
and through a coupling to the trace anomaly of the energy-momentum tensor
---$\theta$ term.
As a result, the following extension of the cross-section for
spin-independent scattering is proposed~\cite{Hoferichter:2016nvd}
\begin{align}
\frac{\text{d} \sigma_{\chi\mathcal{N}}^\text{SI}}{\text{d} \mathbf{q}^2}&=\frac{1}{4\pi\mathbf{v}^2}\Big|c_+^M\mathcal{F}_+^M(\mathbf{q}^2)+c_-^M\mathcal{F}_-^M(\mathbf{q}^2) +c_\pi \mathcal{F}_\pi(\mathbf{q}^2)+c_\pi^\theta \mathcal{F}_\pi^\theta(\mathbf{q}^2)\Big|^2,
\label{cross_section_short}
\end{align}
with each $c$ coefficient sensitive to a different
corner of the parameter space of new-physics models.

The dashed lines of figure~\ref{f:structure_factors}
take into account the interferences with the dominant term.
When these are included, two additional contributions appear:
radius, or momentum-dependent, corrections to the leading operator
---that nevertheless probe a different combination of new-physics parameters---
and the partially coherent operator $\mathcal{O}_3$,
which turns out to give the leading correction among all the
one-nucleon operators proposed in the nonrelativistic EFT.
A further generalization of the coherent cross-section
is therefore suggested~\cite{Hoferichter:2016nvd}:
\begin{align}
\label{cross_section_dec}
\frac{\text{d} \sigma_{\chi\mathcal{N}}^\text{SI}}{\text{d} \mathbf{q}^2}=\frac{1}{4\pi\mathbf{v}^2}&\bigg|\Big(c_+^M-\frac{\mathbf{q}^2}{m_N^2} \, \dot c_+^M\Big)\mathcal{F}_+^M(\mathbf{q}^2)+c_\pi \mathcal{F}_\pi(\mathbf{q}^2)
+c_\pi^\theta \mathcal{F}_\pi^\theta(\mathbf{q}^2)\notag\\
&+\Big(c_-^M-\frac{\mathbf{q}^2}{m_N^2} \, \dot c_-^M\Big)\mathcal{F}_-^M(\mathbf{q}^2)
+\frac{\mathbf{q}^2}{2m_N^2}\Big[c_+^{\Phi''}\mathcal{F}_+^{\Phi''}(\mathbf{q}^2)+c_-^{\Phi''}\mathcal{F}_-^{\Phi''}(\mathbf{q}^2)\Big]\bigg|^2.
\end{align}

Note that not all terms are independent,
as for instance for a Majorana (Dirac) spin $1/2$ WIMP
there are only 4 (7) independent Wilson coefficients,
so that a correlated analysis of several experiments would be required.
A more practical analysis with data from a single experiment
and taking limits on one operator at a time
---for instance on Eq.~\eqref{cross_section_short}---
should take this carefully into account.

\subsection{Spin-dependent and inelastic scattering}

Coherent ---spin-independent--- scattering is not very sensitive to the
detailed nuclear structure of the target nuclei.
This is because at vanishing momentum-transfer,
the structure factor of the leading term is
$\mathcal{F}^M_+ =A^2$,
simply counting the number of nucleons.
The momentum-transfer dependence is given by the nuclear density
\cite{Vietze:2014vsa}.
In contrast, a careful nuclear structure calculation is needed
for spin-dependent scattering, which is very sensitive to
the nuclear spin distribution among all nucleons~\cite{Klos:2013rwa}.

Only odd-$A$ nuclei are sensitive to spin-dependent interactions:
stable even-$A$ nuclei have spin zero due to nuclear pairing.
Therefore, for a given nucleus this interaction is mostly sensitive
to the nucleon species with an odd number of components,
either protons or neutrons.
For odd-$A$ xenon isotopes $^{129}$Xe and  $^{131}$Xe,
with atomic number $Z=54$, neutrons carry most of the spin
and the so-called "structure factor for neutrons"
is orders of magnitude larger than the "structure factor for protons".
This separation is actually not general
as it simply refers to different combinations of isoscalar
---same for neutrons and protons--- or isovector
---opposite--- couplings.
When considering only one-nucleon operators, however,
the separation is valid at vanishing momentum transfer~\cite{Engel:1992bf}:
\begin{align}
\mathcal{F}_{\text{1b}}^{\text{SD}}(q=0)\propto
\bigl|(a_0+a_1)\langle{\bf S}_p\rangle+(a_0-a_1)\langle {\bf S}_n\rangle\big|^2,
\end{align}
with $a_{0/1}$ the isoscalar/isovector couplings
and $\langle{\bf S}_{p/n}\rangle$ the proton/nucleon spin expectation values. 

Two-body currents prevent the validity of the separation.
As illustrated in figure~\ref{f:2b_long}, two-nucleon interactions
do not distinguish between neutrons and protons,
and therefore it is not possible to disentangle proton and neutron
contributions. The structure factor can be generalized as
\cite{Menendez:2012tm}
\begin{align}
\mathcal{F}_{\text{1b+2b}}^{\text{SD}}(q=0)\propto
\bigl|(a_0+a_1[1+\delta])\langle {\bf S}_p\rangle
+(a_0-a_1[1+\delta])\langle {\bf S}_n\rangle\bigr|^2,
\label{eq:sf_sd}
\end{align}
where $\delta\sim-0.2$~\cite{Klos:2013rwa} encodes the two-nucleon contributions
and can be calculated with chiral EFT.
As a result, with two-body currents,
the so-called "structure factor for protons"
---defined by $a_0=-a_1$ in Eq.~\eqref{eq:sf_sd}---
is also sensitive to neutrons,
and increases by over an order of magnitude
with respect to the one-nucleon case because for xenon isotopes
$\langle{\bf S}_{n}\rangle \gg \langle{\bf S}_{n}\rangle$.
This has important practical consequences,
because it makes exclusion limits obtained in experiments using xenon
---which is more sensitive to neutrons--- competitive in "proton cross-sections"
with searches using target nuclei with odd number of protons
---thus more sensitive to protons--- such as fluorine
\cite{Akerib_sd16}.

Once dark matter has been detected, it is left to address
the nature of the dark matter-nucleus interaction.
Spin-dependent scattering can be useful in this respect,
because it could be observed in the elastic and inelastic channels.
The experimental inelastic signature is distinct from the elastic one
---the nucleus $\gamma$ decays to the ground state---
and can be realized if the target nuclei has low-lying excited nuclear states,
such as the 40~keV and 80~keV first excited states in $^{129}$Xe and $^{131}$Xe
~\cite{Baudis:2013bba,McCabe:2015eia}.
For coherent scattering, the inelastic channel is always suppressed
by a factor $A^2$ with respect to the elastic channel~\cite{Vietze:2014vsa},
making it presently undetectable in practice.
Therefore, the observation of an inelastic signal
would clearly point out to a spin-dependent interaction.

\section{Summary}

The most impressive experimental efforts are being made
to unveil the nature of neutrinos and dark matter
in low-energy experiments using nuclei as a source or target.
To make the most of these searches,
comparable theoretical efforts are needed to understand
the nuclear physics driving these processes.
Neutrinoless $\beta\beta$ decay nuclear matrix element calculations differ,
but the most sensitive nuclear structure correlations for the decay
have been identified,
and calculations in larger configuration spaces are underway.
The effect of the weak interaction involving two nucleons
can also be significant,
and explain part of the so-called "$g_A$ quenching".
Ab initio calculations in lighter systems can be performed
to fully understand this "quenching".
Analyses of dark matter searches should consider
all possible interactions of WIMPs with nuclei.
In particular, the coupling to two nucleons can have significant impact
in both coherent and spin-dependent scattering.
The observation of inelastic scattering is a promising way
to determine the nature of the dark matter interaction with nuclei.

%
%
%

\section*{Acknowledgements}
I would like to thank my collaborators
J. Engel, D. Gazit, N. Hinohara, M. Hoferichter, Y. Iwata,
P. Klos, G. Mart{\'i}nez-Pinedo, T. Otsuka, A. Poves, T. R. Rodr{\'i}guez,
N. Shimizu, A. Schwenk, Y. Utsuno and L. Vietze
for very enlightening discussions
and for making use of our common work for these proceedings.
This work has been supported by an International Research Fellowship
the Japan Society for the Promotion of Science (JSPS)
and JSPS Grant-in-Aid for Scientific Research No.\ 26$\cdot$04323.

\end{document}